%% file: jing.tex
\documentstyle[12pt,aaspp4,psfig]{article}

\begin{document}
 
\title{Towards a Unified Model for the `Diffuse Ionized Medium' in Normal
and Starburst Galaxies}
 
\author{Jing Wang$^{1,3}$, Timothy M. Heckman$^{1,3,4}$, and Matthew D.
Lehnert$^{2,3}$}
 
\parindent=0em
\vspace{5cm}
 
1. Department of Physics and Astronomy, The Johns Hopkins University,
Baltimore, MD 21218
 
2. Leiden Observatory, Postbus 9513, 2300RA, Leiden, The Netherlands
 
3. Visiting observers at the Kitt Peak National Observatory of the
National Optical Astronomy Observatories, operated by AURA under
contract with the National Science Foundation.
 
4. Adjunct Astronomer, Space Telescope Science Institute.
 
\parindent=2em
 
\begin{abstract}
The `Diffuse Ionized Medium' (DIM) comprises a significant fraction
of the mass and ionization requirements of the ISM of the Milky Way
and is now known to be an energetically significant component in most
normal star-forming galaxies. Observations of the ionized gas in starburst
galaxies have revealed the presence of gas with striking similarities
to the DIM in normal galaxies: relatively low-surface-brightness and
strong emission from low-ionization forbidden lines like 
[SII]$\lambda\lambda$6716,6731. In this paper we analyze 
H$\alpha$ images and long-slit spectra
of samples of normal and starburst galaxies
to better understand the nature of this diffuse, low-surface-brightness
gas. We find that in both samples there is a strong inverse correlation
between the H$\alpha$ surface-brightness ($\Sigma_{H\alpha}$) 
and the [SII]/H$\alpha$ line ratio at a given location in the galaxy. However,
the correlation for the starbursts is offset brightward
by an order-of-magnitude in H$\alpha$ surface-brightness at a given line ratio.
In contrast, we find that all the galaxies (starburst and normal alike)
define a {\it universal} relation between line ratio and the {\it relative}
H$\alpha$ surface
brightness ($\Sigma_{H\alpha}$/$\Sigma_e$, where $\Sigma_e$ is the
mean H$\alpha$ surface brightness within the galaxy half-light radius).
We show that such a universal correlation is a natural outcome of a model
in which the DIM is photoionized gas that has a characteristic 
thermal pressure ($P$) that is proportional to the mean rate of star-formation
per unit area in the galaxy ($\Sigma_{SFR}$).
Good quantitative agreement with the data
follows if we require the constant of proportionality to be consistent
with the values of $P$ and $\Sigma_{SFR}$ in the local disk of the
Milky Way. Such a scaling between $P$ and $\Sigma_{SFR}$ may arise either
 because feedback from massive stars heats the ISM or because
$\Sigma_{SFR}$ is determined (or limited) by the mean gas pressure.
\end{abstract}


\section{INTRODUCTION}

The `Diffuse Ionized Medium' (DIM) is recognized
as an important component of the ISM in galaxies.
This gas, first discovered as
the `Reynolds Layer' in our Galaxy (see, e.g. Reynolds 1990 for a review),
seems to be ubiquitous in late-type spiral (e.g. Hoopes et al. 1996;
Wang, Heckman \& Lehnert 1997--hereafter WHL) 
and irregular galaxies (e.g. Martin 1997).
Indeed, the universal existence of the DIM has been inferred from
integrated emission-line ratios of a large sample of normal late-type
galaxies (Lehnert \& Heckman 1994).
The observed properties of the DIM are characterized by relatively 
strong low ionization forbidden lines compared to normal HII regions,
a low surface brightness, a rough spatial correlation with HII regions,
and a significant contribution (
$\sim$20\%--40\%) to the global H$\alpha$ luminosity.
 
The observations raise many interesting questions about the physical
and dynamical state of the DIM that are still to be answered.
The energy required to power the DIM suggests that the gas either
soaks up nearly 100\% of the mechanical energy supplied by supernovae
and stellar winds, or the topology of the interstellar medium must
allow roughly 1/3 of the ionizing radiation produced by
massive stars to escape HII regions and propagate into the
disk.
Although the diffuse nature of the DIM suggests that it maintains
pressure balance with the rest of the ISM, little
observational evidence has been collected to support this
idea. Only in the Reynolds layer has the electron density been
derived through observations of
pulsar dispersion measures (e.g. Reynolds 1993) and
seems consistent with the typical ISM thermal pressure of $\sim$3000~K~cm$^{-3}$
(Jenkins et al. 1983). Furthermore, it is still unknown
how or if this pressure is regulated by the hotter coronal-phase gas
created by supernovae.
 
While these questions wait to be answered,
a study of the general properties of the DIM in
galaxies with widely differing star formation rates (SFRs) per unit area
or per unit volume
might provide us with more clues in this
endeavor. The simple reason behind this is that star formation has
a significant impact on the ISM. The energetics and
dynamics of the DIM must therefore be strongly influenced (or even regulated)
by the feedback from star formation (e.g. Elmegreen \& Parravano 1994).
In starbursts, for example,
the intense star formation can provide feedback to
the ISM that might be significantly different {\it qualitatively}
from the case in quiescent galaxies. Dynamically,
the collective effect of supernovae exploding in a hot rarefied medium created
by previous supernovae may minimize radiative losses and thus provide the
energy to
drive a galactic scale outflow in starbursts 
(e.g. Heckman, Armus \& Miley 1990).
Thus, we would expect a more significant
kinematic disturbance in the DIM in starbursts than in normal disks and
a greater role of shock-heating in the energetics
throughout the ISM. That is, the hot coronal-phase gas in starbursts
should be more pervasive than in normal disks (as parameterized by
the {\it porosity}, McKee \& Ostriker 1977) and would therefore have
the potential to regulate
the pressure of the ISM over a relatively larger
 volume. Since the DIM traces the
heating and ionization of gas occupying a substantial fraction
of the volume of the ISM and comprising much of its mass, the above issues are
closely tied to
our understanding of the DIM.
 
We note that among the galaxies that have been searched for a DIM,
most are normal spirals and irregulars. However, observational data
suggest that there is a similar DIM component in starburst galaxies.
For instance, in the nearby starburst NGC 253,
Hoopes et al. (1996) found a faint H$\alpha$-emitting
gas surrounding bright HII regions that is similar to the DIM
in other quiescent spirals. More generally, the spectra
of extra-nuclear regions of starbursts (e.g. Lehnert \& Heckman 1995) show that
the relative strengths
of the low-ionization lines are high---very 
similar to what has been observed in
the DIM in normal galaxies. It is therefore worthwhile to
systematically study the faint emission-line gas in starbursts and
compare this gas to the DIM in normal galaxies.
 
\section{SAMPLE SELECTION AND DATA DESCRIPTION}

In this paper we will explore the emission-line properties
of galaxies with a wide range of star formation rates (SFRs) using
long-slit spectroscopic data. Our sample includes 
both normal, quiescent late-type spirals and IR-selected
starbursts. 
Our data on normal galaxies come from a survey of the DIM
in nearby face-on spirals. This subset involves seven galaxies selected
on the basis of Hubble type (Sb and later), proximity (closer than 10 Mpc),
large angular size ($>$ 10 arcmin), and relatively face-on orientation
(inclination $<$ 65 degrees).
Part of
the data has been presented in Wang et al. (1997), and
the rest can be found in Wang (1998).
The reader can refer to these references for details of the observations
and data reduction.
Complete observational information is summarized in Table 1.

The starbursts are a sample of infrared-bright (S$_{60\micron}$ $>$ 5.4 Jy),
edge-on (a/b $\gtrsim$ 2), infrared-warm
(S$_{60\micron}$/S$_{100\micron}$ $>$ 0.4)
galaxies compiled by Lehnert (1993, dissertation, Johns Hopkins University) 
and also described by
Lehnert and Heckman (1995). 
Analysis of both samples of data can help us understand the physical state
and dynamics of the ionized gas over a wide range of star formation rates.

We note that due to different sample selection criteria, the starbursts
are generally much more distant and more inclined than our normal spirals.
Otherwise, the major difference between the starbursts and the normal
spirals is just the much larger star-formation rates per unit area
($\Sigma_{SFR}$) in the former (Lehnert \& Heckman 1996b).

In order to compare our normal galaxies and starbursts, we used
three quantities that have been measured with long-slit spectroscopy
at many locations within each galaxy
for both groups of galaxies. These are the
line ratios represented by [SII]$\lambda\lambda$6716,6731/H$\alpha$,
the H$\alpha$ surface brightnesses, and the H$\alpha$ or [NII]$\lambda$6584
linewidths. The [SII] doublet was selected to represent low ionization lines
rather than [NII] because Nitrogen has a secondary nucleosynthetic origin,
while Sulfur has a primary one. That is, the S/O ratio is independent of 
metallicity, while N/O $\propto$ metallicity (cf. Vila-Costas \& Edmunds 1993).
As shown by WHL this results in systematic differences in the [NII]/H$\alpha$
ratio in the DIM in galaxies of different metallicity - differences that
are not present in the [SII]/H$\alpha$ ratio.

For our normal galaxies, we have multiple slit positions that
spectroscopically sample representative regions in the disks of these galaxies 
(WHL; Wang 1998). We find that the emission-line properties 
in these galaxies vary spatially in a smooth way from the centers of HII
regions out into the surrounding DIM.

Data for the starbursts are in the form of
long-slit spectra centered on the nuclei and oriented along both the minor
and major axes. The spectra have been extracted using 3 pixel spacing
starting from the nuclei and moving outward along the slit. We can then
examine the variations in starbursts from bright
starburst cores to fainter surrounding nebulae, by analogy with our analysis
of the normal galaxies.
The starbursts are generally much more distant than our
normal spirals. Therefore we sample the gas in starbursts on a 
much larger scale, as each spatial resolution element would encompass a
large number of HII regions.

The [NII]$\lambda$6584 linewidth is used to represent kinematics in
starbursts, while either H$\alpha$ linewidth (if EQW(H$\alpha$) $>$ 3$\AA$)
or [NII] linewidth (if EQW(H$\alpha$) $<$ 3$\AA$) is used (see WHL) for
our normal galaxy sample. This difference in the adopted linewidth is 
unlikely to affect our analysis significantly since 
the H$\alpha$ and [NII] linewidths correlate well with one another for the
starburst (Lehnert \& Heckman 1995) sample. 

The number of galaxies we can use in the starburst sample is
limited by the availability of photometric data and other relevant 
information. We use 32 galaxies from this sample that
have been observed spectroscopically
under nearly photometric conditions and hence can
provide us with line ratios, H$\alpha$ surface brightnesses, and
linewidths. We have excluded from this analysis the Circinus galaxy,
whose H$\alpha$ image in Lehnert \& Heckman (1995) is dominated by the
central AGN and NGC 5253, which is a dwarf galaxy unlike the members
of our normal galaxy sample.

In the following analysis we will attempt to normalize the star-formation rate
in a given location of a galaxy by dividing the measured H$\alpha$ surface
brightness at that location by the average H$\alpha$ surface brightness
in the galaxy. To do so, we
define a galaxy effective H$\alpha$
surface brightness 
$\Sigma_e$ as the ratio of half of the total H$\alpha$ flux to
the solid angle subtended by the area within the H$\alpha$ half-light
radius($\pi r_e^2$). 
This is effectively the average surface brightness within the H$\alpha$
half-light radius.
In order to do this scaling we further selected 19 objects 
out of the subsample of 32 galaxies 
for which Lehnert and Heckman (1995) have measured the H$\alpha$ half-light
radii and total H$\alpha$ fluxes. The measured
total H$\alpha$+[NII] flux is corrected to the H$\alpha$ flux
based on the measured [NII]/H$\alpha$ ratio within $2 r_e$. Thus,
$\Sigma_e$ can be estimated for these 19 starburst  
galaxies and used to scale the
observed values of $\Sigma_{H\alpha}$ as measured at different positions in
each galaxy.

All the spectral data in the normal spiral subsample have been properly
calibrated 
for absolute surface brightness. In addition, we have measured 
$r_e$ from the H$\alpha$ images (WHL and Wang et al. 1998 in preparation). 
Therefore all data from this subsample can be
utilized in this analysis. 

We have not attempted to correct the observed
H$\alpha$ surface brightnesses for the effects of internal reddening.
This is likely to be significant. For example, Kennicutt
(1983) estimates an H$\alpha$ extinction of about 1 magnitude for typical
giant HII regions in normal galaxies, and Armus et al (1989) find a
typical value of about 2 magnitudes for IR-selected galaxies.
Note that while an extinction-correction would increase the absolute
values of the H$\alpha$ surface brightnesses (and implied star-formation
rates per unit area) it will only affect the normalized surface-brightnesses
defined above if the extinction varies spatially. We return to these issues
later in the paper.

\section{RESULTS}

The relative strengths of the low-ionization emission-lines in starbursts
correlate
well with H$\alpha$ surface brightness (Figure 1). Data for HII
regions and the DIM in our normal galaxies are also plotted for comparison.
Both sets of data show higher strengths of low-ionization
lines at lower surface brightness, and they
all suggest a strong continuity in physical state 
between the high surface brightness and low surface brightness gas. 
In fact, the correlations can be roughly described as a power-law relation 
between [SII]/H$\alpha$ and $\Sigma_{H\alpha}$ with similar
slopes for both samples. However, there is a noticeable  offset between
the two groups of galaxies: at a given line-ratio, the gas in
the starbursts has 
an average surface brightness that is about an order-of-magnitude higher than
the gas in
the normal galaxies. This
is not surprising as the starbursts have much higher H$\alpha$ surface
brightnesses in general, and correspondingly 
higher $\Sigma_{SFR}$. We will further examine
this systematic difference in surface brightness in the following
paragraphs. We reject some DIM data points near the nucleus of M 81
due to heating processes other than
pure photoionization by young stars in that region (cf. Devereux, Jacoby,
\& Ciardullo 1995, 1996). 

While the line-ratios correlate strongly with surface-brightness, the starburst
sample shows no correlation between linewidth and
$\Sigma_{H\alpha}$, and only a weak correlation between
linewidth and [SII]/H$\alpha$. 
Linewidths of the gas in starbursts are typically a
few hundred kilometers per second, about a factor of 10
larger than those in normal galaxies. The data for the normal galaxies
bear little resemblance to that of the starbursts with respect to
these kinematic relations.
It is likely that kinematics of the emission-line gas are different
in the two types of galaxies for two major reasons. 
For the distant highly-inclined starbursts, the spectra sample a large
number of emission line nebulae across starburst disks even within a 
small aperture, and the lines may be broadened by relative motions
among the nebulae. Supernova-driven superwinds are also responsible
for much of the linewidth broadening (Lehnert and Heckman 1996a), especially
along the minor axis. Thus we conclude that
linewidths do not track well with other emission-line parameters for the
combined galaxy sample. 

In starbursts the continuity between high surface
brightness regions (starburst nuclei) and low surface brightness
(outer) regions, together with
the lack of correlation between kinematics and emission line 
intensity ratios, suggests that the gas is mainly photoionized
instead of mechanically heated. While we have drawn the same conclusion
for the gas in the normal galaxies (WHL), 
Figure 1 shows a more general trend that is applicable to
the gas in starbursts as well. It is 
not surprising that most of line emission results from
photoionization since the ionizing radiation energy from
OB stars is about an order of magnitude higher than the total
kinetic energy released by supernovae and stellar winds
(cf. Leitherer \& Heckman 1995).

To further explore the possibility of photoionization, 
we need to understand the H$\alpha$ surface brightness offset
between the two groups of galaxies in Figure 1, which
is presumably physically related to the much higher
star formation rates per unit area in the IR-selected starbursts.
To test this idea, we adopt the effective surface brightness $\Sigma_e$
defined in section 2 above to scale the observed values of $\Sigma_{H\alpha}$
and then plot the line-ratio
[SII]/H$\alpha$
versus relative (dimensionless) surface brightness $\Sigma_{H\alpha}/\Sigma_e$ 
(Figure 2).
We have excluded the starburst data beyond $2 r_e$ because
those data are relatively noisy and are more likely affected by 
superwind-driven
shock-heating processes in the galaxy halos (Lehnert \& Heckman 1996a).

Figure 2 shows remarkably universal pattern of line-ratio
variation as a function of {\it normalized} surface brightness. The scaling
by $\Sigma_e$ has successfully eliminated the systematic difference
in surface brightness at a given line-ratio between the two groups of galaxies. 
Figure 2 demonstrates the interesting similarities in 
emission-line properties not only
between HII regions in normal spirals and in starburst nuclei, but 
also between the DIM in normal galaxies and the relatively faint
emission-line gas surrounding starburst nuclei. The universal
continuity between high surface brightness and low surface brightness 
gas suggests that these emission-line properties
all vary with a single parameter and therefore
provides further support to the idea of
photoionization as the dominant mechanism. 

Because of the similarity in emission-line properties, we propose
that the fainter emission-line gas in starbursts and the DIM in normal
galaxies has the same physical nature. While it remains to be confirmed
observationally that the low surface brightness gas in the starbursts is
generically 
diffuse rather than from the sum of many faint, discrete HII regions, our
conjecture
is supported by H$\alpha$ imaging results of the nearest starburst galaxies
like M 82, NGC 253, NGC 5253, and NGC 1569 (cf. Lehnert \& Heckman 1995,
Marlowe et al 1995; Martin 1997). In addition, we 
(Wang 1998; Wang et al. 1998) have analyzed our H$\alpha$ images
of the normal galaxies and find
that there is no {\it absolute} H$\alpha$ surface brightness limit that cleanly
separates the DIM from bright HII regions. Instead, we show that defining the
DIM in terms of a normalized surface-brightness analogous to that described
above leads to a natural segregation that is independent of a galaxy's mean
surface brightness in H$\alpha$. This implies that even in a 
starburst galaxy a DIM component would exist, but this diffuse gas
will have a characteristically high 
absolute surface brightness. Since the mean
$\Sigma_{H\alpha}$ seems to be a reasonable dividing point between the DIM
and HII regions, we suggest that for the sample galaxies discussed in this
paper, a crude surface brightness limit is the $\Sigma_{H\alpha}$
averaged within the half-light radius (=$\Sigma_e$). Therefore
$\Sigma_{H\alpha} / \Sigma_e = 1.0$ could be used to isolate the `DIM'
gas in Figure 2.

For simplicity, in the following discussion we will tentatively 
adopt the same acronym DIM for the 
faint gas in the starbursts. We proceed to use a generic photoionization
model to explain the correlation in Figure 2 and then address other
implications of
the data.

\section{DISCUSSION}
\subsection{Photoionization of the DIM}

The observed inverse correlation between the relative strength of the
low-ionization
lines and the H$\alpha$ surface brightness could have a simple physical
explanation
if the emitting gas clouds all have roughly the same density. In this
case, the higher the value of the local intensity of the ionizing radiation
field, the higher the value of the ionization parameter U (defined to be
the ratio of the densities of ionizing photons and electrons within a
photoionized gas cloud). Simple ionization equilibrium arguments show
that U determines the ionization state of the gas, while
recombination means that
the H$\alpha$ surface brightness will be proportional to the
intensity of the ionizing radiation field. Thus, the
proportionality
between $U$ and the H$\alpha$ surface brightness will naturally produce enhanced
relative intensities of low-ionization lines in the faint gas (cf.
Domg\"{o}rgen \& Mathis 1994).

The {\it generalized} relation between line-ratio and
$\Sigma_{H\alpha}/\Sigma_e$ for the DIM in Figure 2
could then be explained {\it if}
there is a direct
proportionality between the average thermal
pressure in the diffuse interstellar medium
and the average star-formation rate per unit area in the galaxy. 
That is,
the ratio of the H$\alpha$ surface brightness at a particular location
compared to the mean value in the galaxy would then be proportional to the local
value of the intensity of the ionizing radiation field divided by a quantity
that is proportional to the thermal pressure and hence the density of the
photoionized cloud  (T $\sim 10^4$ K).
Thus, $\Sigma_{H\alpha}/\Sigma_e$
$\propto$ $U$.

Later
in this section, we will briefly discuss some of the physics that might
lead to such a proportionality. Here we simply remark that there is
good empirical evidence that this proportionality is roughly obeyed when 
extreme
starbursts like M 82 are compared to the disks of normal spirals like the
Milky Way. In the M 82 starburst, the thermal gas pressure is $P/k$
$\sim$10$^7$ K cm$^{-3}$ (Heckman, Armus, \& Miley 1990)
and the star-formation rate per unit area is $\Sigma_{SFR}$
$\sim$ 30 M$_{\odot}$ yr$^{-1}$
kpc$^{-2}$ for a Salpeter IMF extending from 0.1 to 100 M$_{\odot}$
(cf. Kennicutt 1998). Heckman, Armus, \& Miley (1990) and
Lehnert \& Heckman (1996b) show these are typical values for both
parameters in extreme starbursts.
In comparison, in the local Milky Way disk
the thermal gas pressure is $P/k$ $\sim$ 10$^{3.5}$ K cm$^{-3}$ (cf.
Jenkins et al 1983; Reynolds 1993), and $\Sigma_{SFR}$ $\sim$
4 $\times
10^{-3}$ M$_{\odot}$ yr$^{-1}$ kpc$^{-2}$ 
(McKee \& Williams 1997 adjusted to our adopted IMF). Thus, $\Sigma_{SFR}$
is roughly 7000 times greater in M 82 and the pressure is roughly 3000
times greater. This agrees with Lord et al (1996) who estimate that
both the thermal pressure and FUV intensity in M 82 are three-to-four
orders of magnitude higher than in the local ISM.

Let us for the moment then adopt the conjecture that $P \propto 
\Sigma_{SFR}$, and
derive a relation between $\Sigma_{H\alpha}$, $U$, and $n_{e}$ 
utilizing photoionization models. 
Suppose the gas in the DIM is illuminated by an isotropic
ionizing radiation field. Then the one-sided incident ionizing flux 
$\Phi_{Lyc}$ is related to the observed area-averaged  H$\alpha$
surface brightness of the cloud by
\begin{equation}
  \Phi_{Lyc} = 4 \pi \frac{\Sigma_{H\alpha}}{h\nu} \frac{1}{f_{H\alpha}}
\frac{A_{proj}}{A_{tot}} e_{H\alpha}
\end{equation}
(Vogel et al. 1995) where $e_{H\alpha}$ is the ratio of the
intrinsic (extinction-corrected) and observed H$\alpha$ surface brightness and
$f_{H\alpha}$ is the fraction of
recombinations which produce H$\alpha$ photons
(=0.46 for T = 10$^4$ K and Case B recombination). 
The ratio of observed area to total area $A_{proj} / A_{tot}$
is determined by the nebular geometry. It is 1/2 for a slab and 1/4 for a
sphere. We use 1/3 to represent an average case. According to the definition
of U, we can express $U = 4 \Phi_{Lyc} /n_e c$, therefore
\begin{equation}
  \Sigma_{H\alpha} = 5.9\times10^{-14} \left(\frac{n_e}{1\ {\rm cm^{-3}}}\right)\ U\ e_{H\alpha}^{-1}
\ \ \ \ \ {\rm ergs\ s^{-1}\ cm^{-2}\ arcsec^{-2}}
\end{equation}

To compare this to our data, we need to understand how $n_e$ in the DIM is
related to
the mean star-formation-rate per unit area, as measured by $\Sigma_e$.
Since photoionized gas is generically in thermal equlibrium at 
a temperature of roughly 10$^4$ K (e.g. Osterbrock 1989), relating
$n_e$ to $\Sigma_e$ is equivalent to determing the constant of
proportionality in the relation
$P \propto \Sigma_{SFR}$.
To do this, we will adopt a purely empirical approach for the
moment and insist that this constant agree with values for
$P$ and $\Sigma_{SFR}$ in the ISM
of the Milky Way. Later we will explore the possible physical basis
of this. 

In the local disk of the Milky Way, $\Sigma_{SFR}$ $\sim$ 4 $\times
10^{-3}$ M$_{\odot}$ yr$^{-1}$
kpc$^{-2}$ implies an average intrinsic H$\alpha$ surface brightness for
the disk
of 1.2 $\times$ 10$^{-16}$ ergs s$^{-1}$ cm$^{-2}$ arcsec$^{-2}$
(where we have assumed contnuous star-formation with a Salpeter IMF extending
from 0.1 to 100 $M_{\odot}$ - Leitherer \& Heckman 1995). A 
thermal pressure
of $P/k$  $\sim$ 10$^{3.5}$ K cm$^{-3}$ implies n$_e$ $\sim$ 0.16 cm$^{-3}$
in the photoionized gas. Thus, the predicted relation between $\Sigma_e$
and $n_e$ based on our own Galaxy is:
\begin{equation}
  \Sigma_e = 7.4\times10^{-16} \left(\frac{n_e}{1\ {\rm cm^{-3}}}\right)
 e_{H\alpha}^{-1}
\ \ \ \ \ {\rm ergs\ s^{-1}\ cm^{-2}\ arcsec^{-2}}
\end{equation}

Photoionization models of the DIM (Sokolowski 1993) are able to reproduce
the observed emission-line ratios provided that the
cosmically-abundant, refractory elements (i.e. Fe and Si) are largely
locked-up in dust grains, as in the case of diffuse clouds in our own
Galaxy (cf. Savage \& Sembach 1996 and references therein).
The models also better match the data if the radiation field
incident on the DIM has been hardened due to radiative transfer
en route to the DIM (e.g. there is an optical depth of-order unity at the Lyman
edge between the DIM and the O stars). Adopting these `depleted and hardened'
models, we then
estimate an empirical relation
\begin{equation}
\frac {[SII]\lambda \lambda 6716,6731} {H\alpha} = 1.1\times10^{-2}\ U^{-0.58}
\end{equation}
appropriate for the range of the observed DIM lineratios
([SII]/H$\alpha$ $\approx$ 0.3 -- 1.5).
This enables us to relate [SII]/H$\alpha$ approximately to
$\Sigma_{H\alpha}/\Sigma_e$. The ratio of equations [2] and [3] imply
that $\Sigma_{H\alpha}/\Sigma_e$ = 80 U.

Thus, using Equation [4] above, we obtain:
\begin{equation}
  \frac {[SII]\lambda \lambda 6716,6731} {H\alpha} = 0.14\ \left(\frac {\Sigma_{
H\alpha}} {\Sigma_e}\right) ^{-0.58}
\end{equation}
This relation is represented by the solid line in Figure 2.
The data agree reasonably well with the prediction in the faint gas
(e.g. $\Sigma_{H\alpha}/\Sigma_e$ $<$ 1).
The deviation of the data from the prediction at higher surface brightnesses
($\Sigma_{H\alpha}/\Sigma_e$ $>$ 1) will be briefly discussed in
section 4.3 below.

As a `sanity check' we
now estimate the values of n$_e$ in normal disks and starbursts that are
implied by the measured
values of $\Sigma_e$ based on Eq. [3]. One should keep in
mind that $\Sigma_e$ may need to be corrected for inclination, so the
values given here are upper limits, especially for the
starburst sample where the inclination is high.
We infer an average $n_e$ of 0.4 $e_{H\alpha}$ cm$^{-3}$ from
the observed $\Sigma_e$
for the normal galaxies. 
Taking a typical extinction of 1 magnitude for H$\alpha$ (Kennicutt 1983)
we obtain $n_e$ = 1.0 cm$^{-3}$. This is several times larger than
the value $n_e$ $\sim$ 0.16 cm$^{-3}$
for the Reynolds layer (Reynolds 1993). There might be two major reasons
for this discrepancy. Firstly, the H$\alpha$ extinction in the DIM may be less
than the typical HII region value of 1 magnitude. Secondly, simple 
considerations of hydrostatic equilibrium (see below) imply that the
total ISM pressure (e.g. the sum of thermal, turbulent, cosmic ray, and
magnetic pressures)
decreases with galactocentric distance, so that the thermal pressure and
hence electron density in the DIM may be higher in the inner regions
of galaxies.
Now, $n_e$
for the Reynolds layer has been measured in the solar neighborhood
(about 8 kpc from the Galactic center), while $r_e$ in our normal galaxies
is typically 2--4 kpc (see Table 2). 
The electron density for the DIM in starbursts estimated from
$\Sigma_e$ averages $\sim$~24 cm$^{-3}$,
after correcting for two magnitudes of extinction
in H$\alpha$ (Armus, Heckman, \& Miley 1989). This value for $n_e$ is then
considerably higher than that in normal galaxies (as expected).

To summarize, we have shown that a model in which the DIM in both starburst
and normal galaxies
is photoionized
gas whose thermal
pressure is proportional to the mean rate of star-formation per unit
area in the galaxy can quantitatively
reproduce the observed unified correlation between
the ionization state of the DIM ([SII]$\lambda \lambda$ 6716,6731/H$\alpha$
line ratio) and the relative surface-brightness of the DIM shown in Figure
2. To make this test we have fixed the constant of proportionality
in the relation $P \propto \Sigma_{SFR}$ to its value in the local disk
of the Milky Way.
We now turn to the possible physical basis of this relation.

\subsection{A Supernova-Regulated ISM Pressure}

Suppose we assume that the average thermal gas pressure $P$ within $r_e$
is maintained by the energy and mass released by supernovae (and stellar
winds) inside $r_e$.
We can then relate 
$P$ and therefore $n_e$ in the DIM
to the effective surface brightness $\Sigma_e$. 

Chevalier and Clegg (1985)
have shown that for the case of spherical symmetry and adiabatic conditions
\begin{equation}
 P = 0.12\ \dot{M}^{1/2}\ \dot{E}^{1/2}\ r_e ^{-2}
\end{equation}
Where $\dot{M}$ and $\dot{E}$ are the rates at which gas in the ISM is
shocked and heated respectively.
While this is exact for a spherically-symmetric case, its difference
from a disk geometry can be shown to be negligible (provided that
the gas is adiabatic - see below). 

The starburst models of Leitherer \& Heckman (1995) predict a simple
scaling between the rate at which a starburst would return
kinetic energy ($\dot{E}$) and ionizing photons (Q). Case B
recombination gives the scaling from Q to the H$\alpha$ luminosity.
Now $\dot{M}$ is the amount of mass per unit time
that is heated by supernovae and stellar winds. This will be larger
than the ejecta directly returned from the massive stars by a
`mass-loading' factor $m$. For a standard Salpeter IMF extending up to
100 M$_{\odot}$
and a constant
rate of star-formation for a time longer than 40 Myr, the Leitherer
\& Heckman models and equation 6 above then imply that
$n_e$ is related to $\Sigma_e$ by 
\begin{equation}
  \Sigma_e \simeq 1.4\times10^{-15} \left(\frac{n_e}{1\ {\rm cm^{-3}}}\right)
\ m^{-1/2}\ e_{H\alpha}^{-1}
\ \ \ \ \ {\rm ergs\ s^{-1}\ cm^{-2}\ arcsec^{-2}}
\end{equation}
where we have
assumed ${P} / {2 n_e k} = 10^4$~K. Eq. [7] agrees with the scaling relation
between $\Sigma_e$ and $n_e$ (eq. [3]) of the local disk,  for an appropriate
$m$ value (see below).

We can compare equation [6] to conditions in the local disk of our
own Galaxy.
Based on the rates at which stellar winds and
supernovae inject mass and kinetic energy
($\sim$8$\times$10$^{-4}$ M$\sun$ yr$^{-1}$ kpc$^{-2}$ and
1.2$\times$10$^{39}$ ergs s$^{-1}$ kpc$^{-2}$ respectively) within
3 kpc from the Sun (Abbott 1982; Jura \& Kleinmann 1989), the predicted thermal
pressure of the hot gas is $\sim$2200 m$^{1/2}$ K cm$^{-3}$, compared
to the representative value of $10^{3.5}$ K cm$^{-3}$ from observations of
neutral
(Jenkins et al. 1983) and ionized (Reynolds 1993)
diffuse gas near the Galactic midplane.
The required amount of mass heated per unit time is about twice as much as
that injected directly by supernovae and stellar winds ($m \sim$ 2). Only
about 1/3 of this returned mass comes from high-mass stars
(M$>$ 5 M$_{\odot}$),with the bulk coming from intermediate-mass AGB
stars (Jura \& Kleinmann 1989). The situation in starbursts--where the mass is
returned almost entirely by high-mass stars 
(Leitherer \& Heckman 1995)--is therefore somewhat different. Values for
$m$ $\sim$ 3 to 10 have been estimated in starburst galaxies based on the mass,
luminosity, and temperature of the X-ray emitting gas (e.g. Suchkov et al 1996;
Della Ceca et al 1997; Wang et al 1997).

Equation
[6] assumes that the thermal gas pressure is determined by the deposition
of mass and energy by supernovae and stellar winds, that the hot gas
that results permeates the region of star-formation, and that
radiative losses are negligible. These assumptions
may be valid in starbursts driving superwinds (cf. Heckman,
Lehnert, \& Armus 1993), but probably not in the ISM in normal galaxy disks
where the interaction between stellar ejecta and the ISM is more
complex (cf. Cioffi \& Shull 1991). We therefore consider next
a different physical interpretation of the relation between gas
pressure and star-formation intensity.

\subsection{Hydrostatic Equilibrium and Pressure-Regulated Star Formation}

As discussed above, the results in Figure 2 can be understood
if the DIM in both normal and starburst galaxies
is photoionized, has a roughly constant characteristic density in each galaxy,
and the characteristic thermal pressure in the DIM is proportional to the rate
of star-formation per unit area in that galaxy. In section 4.2, we considered
the possibility that the physical coupling was provided by the energy and
mass deposited in the ISM by supernovae and massive stars.
Here, we consider a different interpretation, namely that the total pressure
in the ISM is specified by hydrostatic equilibrium (e.g. Boulares \& Cox
1990), and that the star-formation rate per unit area is related to, or
perhaps limited by, this
pressure (cf. Dopita 1985).

In a case of
simple hydrostatic
equilibrium, the total (thermal, turbulent, cosmic ray, plus magnetic)
mid-plane pressure in a disk galaxy is given
by
\begin{equation}
  P_{tot} \propto \Sigma_{gas} (\Sigma_{*} + \Sigma_{gas}) \propto \Sigma_{gas} 
\Sigma_{tot}
\end{equation}

On empirical grounds, it is well-established that the star-formation-rate
per unit area ($\Sigma_{SFR}$) in disk galaxies scales with both $\Sigma_{gas}$
and $\Sigma_{tot}$. This suggests that there
might be a simple, direct scaling between $\Sigma_{SFR}$ and $P_{tot}$.
In fact, Dopita \& Ryder (1994) parameterize the problem as
\begin{equation}
\Sigma_{SFR} \propto \Sigma_{gas}^{m} \Sigma_{tot}^{n}
\end{equation}
and find empirically that $m + n = 2.0 \pm 0.5$.
Kennicutt (1998) finds that $\Sigma_{SFR} \propto \Sigma_{gas}^{1.4}$,
while Figure 1 in Dopita \& Ryder (1994) implies 
$\Sigma_{SFR} \propto \Sigma_{*}^{0.6}$. 
Except in the most extreme starbursts, it is reasonable to take
$\Sigma_{tot} \gg \Sigma_{gas}$, so that this last result means roughly that
$\Sigma_{SFR} \propto \Sigma_{tot}^{0.6}$. Combining these results suggests
that:
\begin{equation}
\Sigma_{SFR} \propto \Sigma_{gas}^{1.4} \Sigma_{tot}^{0.6} \propto
P_{tot} (\Sigma_{gas}/\Sigma_{tot})^{0.4}
\end{equation}

Since there is only a small observed variation in $\Sigma_{gas}/\Sigma_{tot}$
(factors of a few) in the disks of late-type galaxies and typical
starbursts, this implies that there should be a proportionality between
$P_{tot}$ and $\Sigma_{SFR}$. If we now assume that the thermal component
of the pressure scales with the total pressure, this is just what we require in
order to understand Figure 2.

As we emphasized in section 4.1 above, the rough quantitative agreement
between photoionization models and the properties of the DIM in the
starbursts and normal galaxies is independent of the nature of the physical,
causal connection between the thermal pressure in the DIM and $\Sigma_{SFR}$.
 The
agreement shown in Figure 2 is based simply on requiring that
the constant of proportionality between these two quantities is consistent
with values in the local disk of our Galaxy.

\subsection{The High Surface-Brightness Gas}

While the model of a photoionized, roughly isobaric DIM provides a
satisfactory quantitative match to the data on the low surface brightness
gas ($\Sigma_{H\alpha}/\Sigma_e$ $<$ 1), Figure 2 shows that there
is a systematic offset between the predictions and the data in the high surface
brightness
range ($\Sigma_{H\alpha}/\Sigma_e$ $\sim$ 1 -- 10). This disagreement is
in the sense that the observed emission-line ratio [SII]/H$\alpha$ is
too high, and therefore that the actual ionization parameter $U$ in the gas
must be smaller than predicted. This could be explained if the density in the
high-surface-brightness gas is higher than estimated in the model.
This is entirely plausible, since the high 
surface-brightness gas (the HII regions) will likely be 
significantly over-pressured
with respect to the surrounding diffuse ISM (e.g. the HII regions may
be self-gravitating or expanding into the lower-pressure DIM).
More quantitatively, we note that the offset between the model and data for
log([SII]/H$\alpha$) in the bright gas
(a difference of $\sim$ 0.6 dex on average) would translate into
a difference of a factor of $\sim$10 lower $U$ and hence higher $n_e$.

An additional factor is that the Sokolowski models we have utilized for the
DIM: 1) adopted a dust-depleted abundance pattern and 2) assumed that
the ionizing radiation field had been hardened as it propagated to the DIM.
Partial depletion onto grains may occur in HII regions (cf. Garnett
et al 1995), but the assumption of spectral hardening is not appropriate
for the HII regions. Dropping these assumptions would decrease the
predicted ratio of [SII]/H$\alpha$ for a given U, and make the
discrepancy worse in Figure 2.
This would require a decrease in U
(and increase in $n_e$) by an additional factor of $\sim$ 3.

Using the value for $n_{e}$ in the DIM in normal galaxies estimated from
$\Sigma_e$
above, we would then require a density of $\sim$30 cm$^{-3}$ in the
HII regions. 
This agrees reasonably well with the
average values of $n_e$ in disk HII regions of
 $\sim$10--100 cm$^{-3}$
(e.g. O'Dell and Casta$\tilde{\rm n}$eda 1984;
Kennicutt, Keel \& Blaha 1989) measured with the [SII] and [OII] doublets.
Our measurements of [SII]$\lambda$6716/[SII]$\lambda$6731 for bright
HII regions suggest similar values for $n_e$.

Following the same reasoning,
we would estimate that the required density in the high-surface
brightness gas in the centers of the starbursts must also be
about 30 times higher than in the low-surface-brightness gas:
$n_e\sim700\ cm^{-3}$.
This is in satisfactory agreement with the directly measured central
densities of
300 to 1000
cm$^{-3}$ (e.g. Heckman, Armus and Miley 1990; Lehnert and Heckman 1996a).

\section{CONCLUSIONS}

We have compared the emission-line properties of the low surface brightness 
gas in starburst galaxies with the DIM in normal spirals. Both samples
show similar attributes of enhanced low-ionization forbidden-line strengths 
(as represented by [SII]/H$\alpha$) relative to typical HII region 
values and a strong inverse correlation between H$\alpha$ surface-brightness
and the [SII]/H$\alpha$ line ratio (in the form of a
smooth transition from high surface brightness 
to low surface brightness regions). The gas kinematics show no strong 
correlation with surface brightness and line ratio in the combined samples.

Although the H$\alpha$ surface brightness corresponding to a given
[SII]/H$\alpha$ line ratio
is preferentially about an order-of-magnitude
larger in starbursts than in normal galaxies, 
we have demonstrated that this can be understood as a consequence of the
proportionately higher mean 
H$\alpha$ surface brightnesses of the starbursts. That is, we have shown that
the {\it relative} surface 
brightness at a particular location, defined as the absolute surface
brightness there ($\Sigma_{H\alpha}$) scaled by 
the mean surface brightness within the H$\alpha$ half-light radius
($\Sigma_e$) for the galaxy as-a-whole,
exhibits a remarkably universal correlation with the
[SII]/H$\alpha$ line ratio for normal and starburst galaxies alike.
This suggests that the emission-line 
properties of the low surface brightness gas in both groups of galaxies 
can be unified to a simple relation between line ratio and relative surface 
brightness, and that the variations in line ratio and relative
surface-brightness are controlled by a 
single parameter.

We have constructed a simple photoionization model to explain 
the correlation between $\Sigma_{H\alpha}$/$\Sigma_e$ and line ratio.
We have pointed out that the [SII]/H$\alpha$ line ratio has an inverse
dependence
on the ionization parameter $U$ (the local ratio of ionizing photons
and electrons in the photoionized gas). For simple recombination,
$\Sigma_{H\alpha}$ is
proportional to the local intensity of the ionizing radiation field. {\it If}
the average thermal pressure in the diffuse ISM in a galaxy ($P$) is
 proportional to
the average rate of star-formation per unit area ($\Sigma_{SFR}$), then
since $\Sigma_{SFR}$ can be measured by $\Sigma_e$,
it follows that $U \propto
\Sigma_{H\alpha}/\Sigma_e$. We have argued that this result naturally
explains the universal dependence of line ratios on relative
surface brightness.

Our simple model is able to quantitatively reproduce the data for normal
and starburst galaxies provided that the constant of proportionality
in the relation between $P$ and $\Sigma_{SFR}$ is consistent with the observed
values for both quantities in local Galactic ISM. Thus, we 
have emphasized that the agreement between our simple
photoionization model and the data is independent of the detailed
physical connection between star formation and ISM pressure.

We have discussed two ways in which $P$
might be physically related to $\Sigma_{SFR}$.
Following Chevalier \& Clegg (1985) we have first 
assumed that $P$ is regulated by the feedback of mass and energy
from supernovae and massive stars. Scaling the amount of mass heated
per supernova so that the predicted thermal pressure matches the
observed pressure in the local Milky Way disk, we found that the
photoionization model agrees roughly with the DIM 
data in both starburst and normal galaxies. 
As an alternative, we explored the possibility (e.g. Dopita 1985) that
$\Sigma_{SFR}$ is determined (or limited) by the total (thermal, turbulent,
cosmic ray, plus magnetic) pressure
$P_{tot}$, and that 
$P_{tot}$ and $P$ are determined by a simple hydrostatic 
equilibrium condition in galactic disks, i.e.
$P \propto P_{tot} \propto \Sigma_{tot}
\Sigma_{gas}$. We used recent empirical results from Kennicutt (1998) and
Dopita \& Ryder (1994) to argue that
$\Sigma_{SFR} \propto P f_{gas}^{0.4}$, where f$_{gas}$ is the
fractional gas mass in the disk. Since f$_{gas}$ varies only by small factors,
this means that $\Sigma_{SFR}$ does roughly scale with $P$. 

The simple model can not account for the emission-line ratios in the high
surface-brightness gas (the giant HII regions)
unless the densities and thermal pressures there are roughly 30
times larger than in the DIM. We argue that this is both reasonable physically
and in agreement with measurements.

We conclude that the low surface brightness gas in the
starbursts shares a common nature with the DIM in the normal galaxies, and
propose that the former can be regarded as the same gas phase as the latter.
Further morphological observations of the low surface brightness gas in
starbursts can confirm this suggestion.

\acknowledgments
 
  We thank S. Baum, D. Calzetti,
R. Kennicutt, R. Wyse, C. Norman, C. Martin, and A. Ferguson
for useful discussions and an anonymous referee for constructive suggestions.

\clearpage

\begin{figure}
\caption{ The [SII]$\lambda\lambda$6717,6731/H$\alpha$ line ratio vs.
the H$\alpha$ surface brightness at the same position in the galaxy.
The starburst data are represented by
small triangles and include 32 galaxies for which photometric 
measurements are available (Lehnert \& Heckman 1995).
These data cover the regions from starburst
nuclei to their outer disks. The data for normal galaxies
are represented by different symbols as follows.
Stars---M 101, triangles---M 51, solid circles---M 81,
crosses---NGC 4395, open circles---NGC 2403, diamonds---NGC 6946,
`I'---IC 342. 
These data are from the DIM, bright HII regions, and intermediate
regions around isolated HII regions. 
The DIM points lie to the upper
left, the HII regions to the lower right, and the intermediate points
lie in between.}
\end{figure}

\begin{figure}
\caption{ The [SII]$\lambda\lambda$6717,6731/H$\alpha$ line ratio vs. the
relative H$\alpha$ surface brightness $\Sigma_{H\alpha}/\Sigma_e$ where
$\Sigma_e$ is the mean surface brightness within the 
H$\alpha$ half-light radius r$_e$.
Symbols are the same as Figure 1. The data includes only 19 starburst galaxies 
(in the Lehnert \& Heckman 1995 sample) that have measured values for
$\Sigma_e$. Regions beyond $2 r_e$ are excluded because the measurement errors
are large and shock-heating by superwinds may be significant in these outer
regions.
This figure demonstrates the universal
relation between line-ratio and relative surface brightness of the
diffuse emission-line
gas for galaxies
with different SFRs. The solid line is the prediction from a simple
photoionization model for the DIM
assuming the thermal pressure in the diffuse gas is related to the
 star-formation rate
per unit area (see text for details). The assumptions in the photoionization
models are not appropriate for the high-surface-brightness 
gas---the HII regions with $\Sigma_{H\alpha}/\Sigma_e$ $>$ 1.}
\end{figure}

\clearpage
\include{jing_tab1}
\include{jing_tab2a}
\include{jing_tab2b}

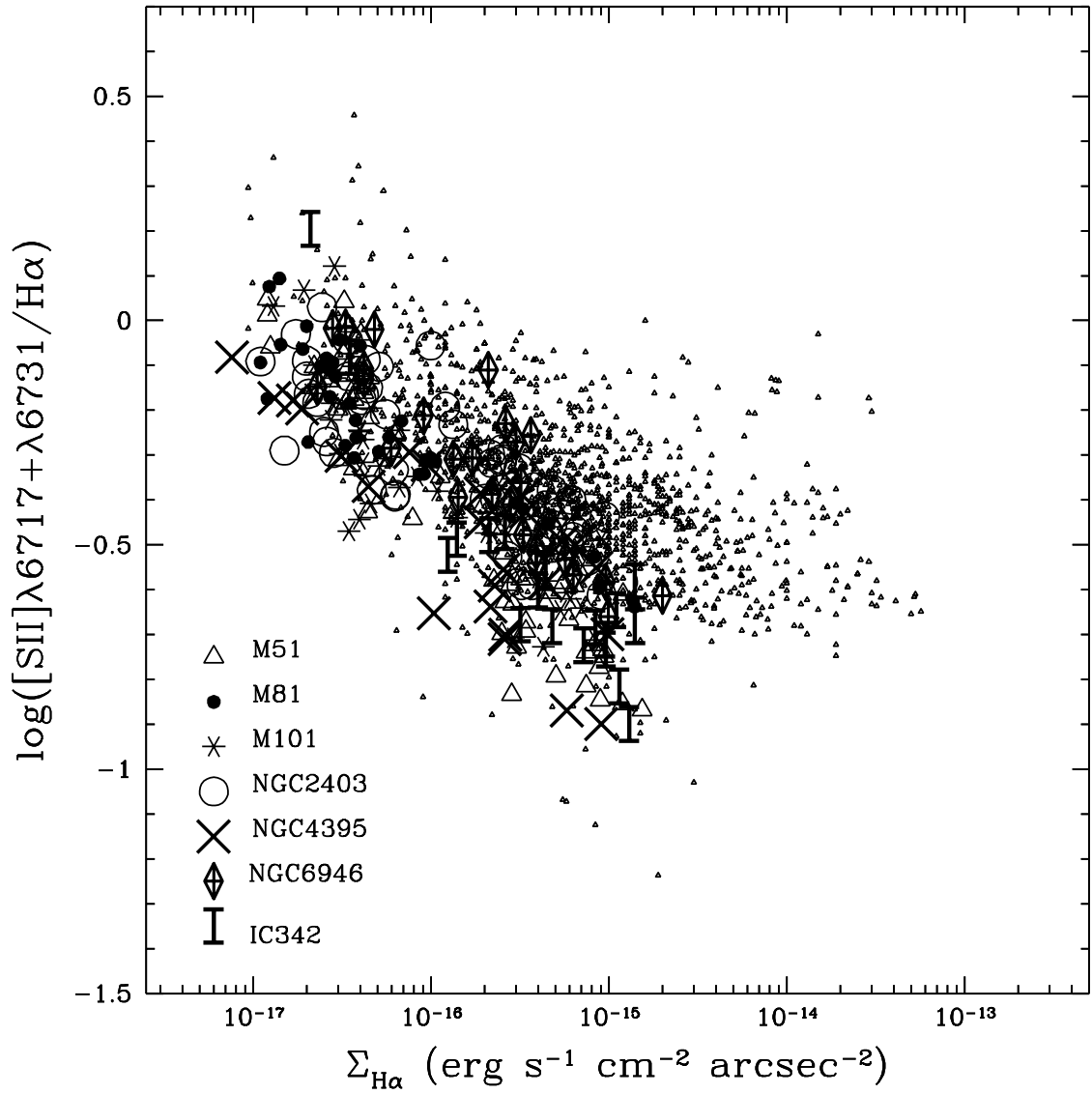
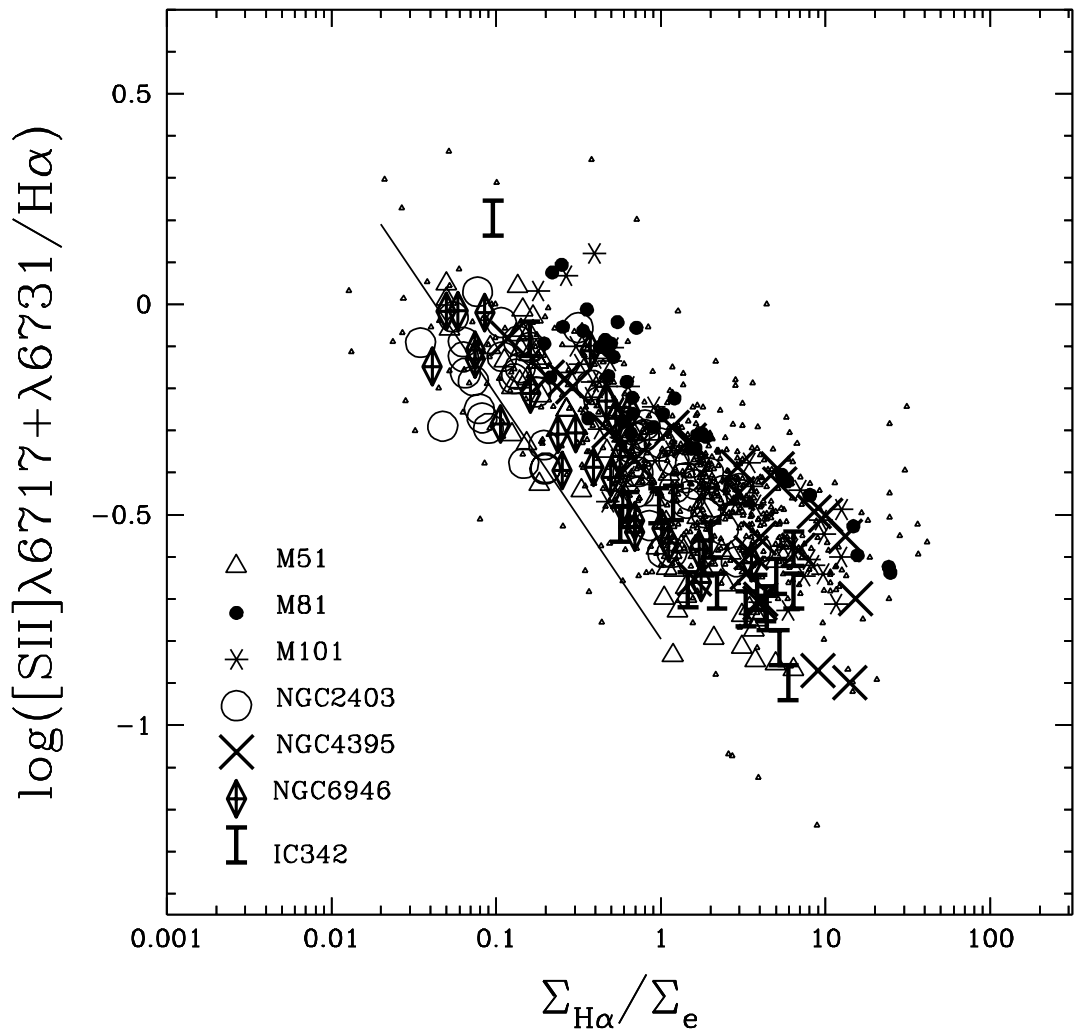
\end{document}

%% file: jing_tab1.tex

\begin{deluxetable}{lcclccccc}
\tablewidth{0pt}
\scriptsize
\tablecaption{High-Resolution Spectroscopic Observations\tablenotemark{a}}
\tablehead{
\colhead{Position\tablenotemark{b}}      &  \colhead{$\lambda_{c}$} & 
\colhead{Coverage\tablenotemark{c}} & \colhead{Grating\tablenotemark{d}} & 
\colhead{Resolution} & 
\colhead{PA} &
\colhead{Blocking Filter} & \colhead{Standard Star} & \colhead{Date} \\
\colhead{} & \colhead{(\AA)} & \colhead{(\AA)} & \colhead{} & 
\colhead{(\AA)} & \colhead{($\arcdeg$)} & \colhead{} & \colhead{} & \colhead{} }

\startdata

M 51: & & & & & & & & \nl
 nuc     &  6563 & 6180-6950 & KPC-24  & 0.8-1.0 &  60 & GG495 & BD +262606
& 02/17/96 \nl
 nuc    &  4850 & 4450-5250 & KPC-24  & 1.2-1.5 & 60 & BG39 & Feige 34 
        & 12/04/96 \nl
 92\arcsec\,N\ 40\arcsec\,W & 6500 & 6120-6880 & KPC-24  & 0.9-1.1 & 60
        & GG496 & Feige 34 & 01/31/97 \nl
 92\arcsec\,N\ 40\arcsec\,W & 4900 & 4500-5300 & KPC-24  & 1.0-1.2 & 60
        & 4-96 & HZ 44 & 01/31/97 \nl
 92\arcsec\,N\ 40\arcsec\,W & 4900 & 4500-5300 & KPC-24  & 1.0-1.2 & 60 
        & 4-96 & G191B2B & 02/03/97 \nl
 115\arcsec\,S\ 20\arcsec\,E & 6563 & 6180-6950 & KPC-24  & 0.8-1.0 & 60
        & GG495 & BD +262606 & 02/18/96 \nl
\tableline
M 81: & & & & & & & & \nl
 nuc    &  6563 & 6180-6950 & KPC-24  & 0.8-1.0 &  60 & GG495 & BD +262606
        & 02/17/96 \nl
 nuc    &  4800 & 4360-5230 & KPC-18C & 1.2-1.5 & 60 & GG385+CUSO4 & 
        Feige 34 & 12/02/96 \nl
 316\arcsec\,S\ 54\arcsec\,E & 6600 & 6220-6980 & KPC-24 & 0.8-1.0 & 60 
        & GG495 & Feige 34 & 12/03/96 \nl
 316\arcsec\,S\ 54\arcsec\,E & 4850 & 4450-5250 & KPC-24  & 1.2-1.5 & 60 
        & BG39 & Feige 34 & 12/04/96 \nl
 234\arcsec\,N\ 183\arcsec\,W & 6500 & 6120-6880 & KPC-24  & 0.9-1.1 & 60
        & GG496 & Feige 34 & 01/31/97 \nl
 234\arcsec\,N\ 183\arcsec\,W & 4900 & 4500-5300 & KPC-24  & 1.0-1.2 & 60 
        & 4-96 & G191B2B & 02/03/97 \nl
\tableline
M 101: & & & & & & & &  \nl
 nuc    &  6563 & 6180-6950 & KPC-24 & 0.8-1.0 & 140 & GG495 & BD +262606 & 02/17/96 \nl
 nuc    &  4900 & 4500-5300 & KPC-24 & 1.0-1.2 & 140 & 4-96 & HZ 44 & 01/31/97 \nl
 286\arcsec\,S\ 202\arcsec\,W & 6500 & 6120-6880 & KPC-24  & 0.8-1.0 & 140 
        & GG495 & HZ 44 & 02/03/97 \nl
 286\arcsec\,S\ 202\arcsec\,W & 4900 & 4500-5300 & KPC-24  & 1.0-1.2 & 140 
        & 4-96 & G191B2B & 02/03/97 \nl
\tableline
NGC 2403: & & & & & & & &  \nl
 nuc    &  6563 & 6180-6950 & KPC-24 & 0.8-1.0 &  35 & GG495 & BD +262606
        & 02/17/96 \nl
 nuc    &  4800 & 4360-5230 & KPC-18C & 1.2-1.5 & 35 & GG385+CUSO4 & 
        Feige 34 & 12/02/96 \nl
 nuc    &  4900 & 4500-5300 & KPC-24 & 1.0-1.2 & 35 & 4-96 & G191B2B & 02/03/97        \nl
 57\arcsec\,S\ 77\arcsec\,E & 6600 & 6220-6980 & KPC-24 & 0.8-1.0 & 35 & GG495 
        & Feige 34 & 12/03/96 \nl
 57\arcsec\,S\ 77\arcsec\,E & 4850 & 4450-5250 & KPC-24  & 1.2-1.5 & 35 
        & BG39 & Feige 34 & 12/04/96 \nl
 57\arcsec\,S\ 77\arcsec\,E & 4900 & 4500-5300 & KPC-24  & 1.0-1.2 & 35 
        & 4-96 & G191B2B & 02/03/97 \nl
 136\arcsec\,S\ 180\arcsec\,E & 6500 & 6120-6880 & KPC-24  & 0.9-1.1 & 35
        & GG496 & Feige 34 & 01/31/97 \nl
\tableline
NGC 4395: & & & & & & & &  \nl
 nuc    &  6563 & 6180-6950 & KPC-24 & 0.8-1.0 &  60 & GG495 & BD +262606
        & 02/17/96 \nl
 nuc    &  4800 & 4360-5230 & KPC-18C & 1.2-1.5 & 60 & GG385+CUSO4 
        & Feige 34 & 12/02/96 \nl
 74\arcsec\,N\ 66\arcsec\,W & 6600 & 6220-6980 & KPC-24 & 0.8-1.0 & 60 & GG495 
        & Feige 34 & 12/03/96 \nl
 74\arcsec\,N\ 66\arcsec\,W & 4850 & 4450-5250 & KPC-24  & 1.2-1.5 & 60 
        & BG39 & Feige 34 & 12/04/96 \nl
 109\arcsec\,S\ 101\arcsec\,E & 6563 & 6180-6950 & KPC-24  & 0.8-1.0 & 60
        & GG495 & BD +262606 & 02/18/96 \nl
\tableline
NGC 6946: & & & & & & & &  \nl
 nuc & 6600 & 6220-6980 & KPC-24 & 0.8-1.0 & 140 & GG495 & Feige 34 & 12/03/96 \nl
 nuc & 4800 & 4360-5230 & KPC-18C & 1.2-1.5 & 140 & GG385+CUSO4 & Feige 34 & 12/02/96 \nl
  35\arcsec\,S\ 175\arcsec\,W & 6600 & 6220-6980 & KPC-24 & 0.8-1.0 & 140 
        & GG495 & BD 284211 & 12/04/96 \nl
\tableline
IC 342: & & & & & & & &  \nl
 nuc & 6600 & 6220-6980 & KPC-24 & 0.8-1.0 & 150 & GG495 & Feige 34 & 12/03/96 \nl
 nuc & 4800 & 4360-5230 & KPC-18C & 1.2-1.5 & 150 & GG385+CUSO4 & Feige 34 & 12/02/96 \nl
 286\arcsec\,N\ 84\arcsec\,E & 6600 & 6220-6980 & KPC-24 & 0.8-1.0 & 150 & GG495 & Feige 34 & 12/03/96 \nl
 286\arcsec\,N\ 84\arcsec\,E & 4800 & 4360-5230 & KPC-18C & 1.2-1.5 & 150 & GG385+CUSO4 & Feige 34 & 12/02/96 \nl
\tableline 

\tablenotetext{a}{All observations were made with the KPNO 4m telescope using
the RC spectrograph and the T2KB CCD. 
A slit of 5\arcmin\ long and 2\arcsec\ wide is used 
centered on either the nucleus or the indicated offsets from the nucleus
of each object. }
\tablenotetext{b}{Centers of the slit. Numbers represents offsets from the
galactic nucleus. }
\tablenotetext{c}{Useful spectral range estimated from 1500 pixels of the CCD. }
\tablenotetext{d}{All gratings are set in second order.}
\enddata

\end{deluxetable}


%% file: jing_tab2a.tex

\begin{deluxetable}{lcccc}
\tablecaption{Effective H$\alpha$ Surface Brightness}
\tablenum{2a}
\tablehead{
\colhead{Object}   &
\colhead{L$_{{\tt H}\alpha}$\tablenotemark{a} } & 
\colhead{r$_{e}$\tablenotemark{b}} &
\colhead{D\tablenotemark{c}} &
\colhead{$\Sigma_{e}$\tablenotemark{d}} \\
\colhead{} & \colhead{(L$\sun$)} & 
\colhead{(kpc)} & 
\colhead{(Mpc)} & 
\colhead{(ergs s$^{-1}$ cm$^{-2}$ arcsec$^{-2}$)}
}
\startdata
M 51  &    4.4$\times$10$^{7}$  &  4.2 & 8.4  & 2.4$\times$10$^{-16}$  \nl
M 81  &    1.5$\times$10$^{7}$  &  4.2 & 3.6  & 6.0$\times$10$^{-17}$    \nl
M 101 &    6.3$\times$10$^{7}$  &  10.2 & 7.4  & 7.0$\times$10$^{-17}$   \nl
NGC 2403 &  1.4$\times$10$^{7}$  &  2.1  &  3.2 & 3.2$\times$10$^{-16}$  \nl
NGC 4395 &  1.9$\times$10$^{6}$  &  1.7  &  2.6  &  6.0$\times$10$^{-17}$  \nl
NGC 6946 &  1.2$\times$10$^{8}$  &  4.5  &  5.9  &  5.6$\times$10$^{-16}$ \nl
IC 342  &   2.3$\times$10$^{7}$  &  2.9  &  2.1  &  2.2$\times$10$^{-16}$ \nl

\tablenotetext{a}{Integrated H$\alpha$ luminosity corrected  for foreground 
  Galactic extinction. Data are from WHL and Wang et al. 1998 (in preparation).}
\tablenotetext{b}{Half-light radius in H$\alpha$ measured at an inner circle
  in the 
  images that enclosed one-half of the total H$\alpha$ light from the galaxy. }
\tablenotetext{c} {Distance adopted from WHL and Wang et al. 1998 
  (in preparation).}
\tablenotetext{d}{Mean H$\alpha$ surface brightness within half-light radius
  r$_{e}$ derived from L$_{{\tt H}\alpha}$ (Col. 2) and r$_{e}$ (Col. 3). 
  The values have been corrected 
  for foreground Galactic extinction.
  The values can be converted to emission
  measure based on the relation that
  5$\times$10$^{-17}$ erg s$^{-1}$ cm$^{-2}$ arcsec$^{-2}$ 
  corresponds to 25 pc cm$^{-6}$, 
  assuming an electron temperature of 10$^4$ K. }

\enddata
\end{deluxetable}


%% file: jing_tab2b.tex

\begin{deluxetable}{lcccc}
\scriptsize
\tablecaption{Effective H$\alpha$ Surface Brightness}
\tablenum{2b}
\tablehead{
\colhead{Object}   &
\colhead{L$_{{\tt H}\alpha}$\tablenotemark{a} } & 
\colhead{r$_{e}$\tablenotemark{b}} &
\colhead{D\tablenotemark{c}} &
\colhead{$\Sigma_{e}$\tablenotemark{d}} \\
\colhead{} & \colhead{(L$\sun$)} & 
\colhead{(kpc)} & 
\colhead{(Mpc)} & 
\colhead{(ergs s$^{-1}$ cm$^{-2}$ arcsec$^{-2}$)}
}
\startdata
NGC1134      &   6.9$\times$10$^7$     &  3.7     &  46.0  &  6.0$\times$10$^{-16}$  \nl
IR03359+1523 &   1.6$\times$10$^8$    &  1.4      &  141.7  &   1.0$\times$10$^{-14}$  \nl
NGC1511      &   7.6$\times$10$^6$     &  0.98     &  18.0  &   9.6$\times$10$^{-16}$  \nl
NGC1572      &   4.0$\times$10$^7$     &  3.0      &  80.5  &   5.4$\times$10$^{-16}$  \nl
NGC1808      &   1.2$\times$10$^7$     &  0.32     &  13.6  &   1.4$\times$10$^{-14}$  \nl
NGC2966      &   1.6$\times$10$^7$     &  0.35     &  31.7  &   1.6$\times$10$^{-14}$  \nl
IC564        &   5.6$\times$10$^7$     &  5.6      &  85.6  &   2.1$\times$10$^{-16}$  \nl
NGC3044      &   1.4$\times$10$^7$     &  1.9      &  21.7  &  4.5$\times$10$^{-16}$  \nl
NGC3511      &   1.1$\times$10$^7$     &  2.3      &  19.4  &   2.5$\times$10$^{-16}$  \nl
NGC3593      &   4.1$\times$10$^6$     &  0.49     &  12.9  &    2.0$\times$10$^{-15}$  \nl
UGC6436     &   7.6$\times$10$^7$     &  2.6       &  143.5  &   1.4$\times$10$^{-15}$  \nl
NGC4433     &   6.1$\times$10$^7$     &  2.6       &  44.8  &   1.1$\times$10$^{-15}$  \nl
NGC4527     &   2.9$\times$10$^7$     &  5.9      &  27.7  &   1.0$\times$10$^{-16}$  \nl
NGC4666     &   7.5$\times$10$^7$     &  5.0      &  24.8  &   3.6$\times$10$^{-16}$  \nl
NGC5073     &   1.2$\times$10$^7$     &  3.0      &  40.9  &   1.7$\times$10$^{-16}$  \nl
NGC5104     &   2.5$\times$10$^7$     &  2.6      &  79.2  &   4.4$\times$10$^{-16}$  \nl
NGC5775     &   3.7$\times$10$^7$     &  4.0      &   25.2  &  2.8$\times$10$^{-16}$  \nl
ZW049.057   &   4.6$\times$10$^6$     &  0.71     &   49.7  &  1.1$\times$10$^{-15}$  \nl
IC5179      &   1.0$\times$10$^8$     &  3.3      &   42.3  &  1.1$\times$10$^{-15}$  \nl

\tablenotetext{a}{Integrated H$\alpha$ luminosity corrected  for foreground 
  Galactic extinction. Data are from H$\alpha$ + [NII] luminosities listed by
  Lehnert \& Heckman (1995) and have been corrected to pure H$\alpha$ 
  luminosities based on average [NII]/H$\alpha$ ratios within 2 half-light 
  radii.}
\tablenotetext{b}{Half-light radius in H$\alpha$ measured at an inner circle
  in the 
  images that enclosed one-half of the total H$\alpha$ light from the galaxy 
  (Lehnert \& Heckman 1995). }
\tablenotetext{c}{Distances adopted from Lehnert \& Heckman 1995.}
\tablenotetext{d}{Mean H$\alpha$ surface brightness within half-light radius
  r$_{e}$ derived from L$_{{\tt H}\alpha}$ (Col. 2) and r$_{e}$ (Col. 3). 
  The values have been corrected 
  for foreground Galactic extinction.
  The values can be converted to emission
  measure based on the relation that
  5$\times$10$^{-17}$ erg s$^{-1}$ cm$^{-2}$ arcsec$^{-2}$ 
  corresponds to 25 pc cm$^{-6}$, 
  assuming an electron temperature of 10$^4$ K. }

\enddata
\end{deluxetable}


%% file: jing.bbl
\clearpage
 
\begin{thebibliography}{}

\bibitem[]{} Abbott, D. C. 1982, \apj, 263, 723
\bibitem[]{} Armus, L., Heckman, T. M. \& Miley, G. K. 1989, \apj, 347, 727
\bibitem[]{} Boulares, A. \& Cox, D. P. 1990, \apj, 365, 544
\bibitem[]{} Chevalier, R. A. \& Clegg, A. W. 1985, Nature, 317, 44
\bibitem[]{} Cioffi, D.F. \& Shull, J. M. 1991, \apj, 367, 96
\bibitem[]{} Della Ceca, R., Griffiths, R. E. \& Heckman, T. M. 1997, \apj, 485, 581
\bibitem[Devereux et al. 1995, 1996]{dev95} Devereux, N. A., Jacoby, G. 
  \& Ciardullo, R. 1995, \aj, 110, 1115  \nl
  ----------. 1996, \aj, 111, 2115 ({\em Erratum})   
\bibitem[Domg\"{o}rgen \& Mathis 1994]{dom94} Domg\"{o}rgen, H. \& Mathis, J. S. 1994, \apj, 428, 647
\bibitem[]{} Dopita, M. A. 1985, \apj, 295, L5
\bibitem[]{} Dopita, M. A. \& Ryder, S. D. 1994, \apj, 430, 163
\bibitem[]{} Elmegreen, B. G. \& Parravano, A. 1994, \apj, 435, L121
\bibitem[]{} Garnett, D. R. et al. 1995, \apj, 449, 77
\bibitem[]{} Heckman, T. M., Armus, L. \& Miley, G. K. 1990 \apjs, 74, 833
\bibitem[]{} Heckman, T. M., Lehnert, M. D. \& Armus, L. 1993, in The
  Environment \& Evolution of Galaxies, ed. J. M. Shull \& H. A. Thronson,
  Jr. (Dordrecht:Kluwer), 455
\bibitem[]{} Hoopes, C. G., Walterbos, R. A. M., Greenwalt, B. E. 1996, 
  \aj, 112, 1429
\bibitem[]{} Jenkins, E. B., Jura, M. \& Loewenstein, M. 1983, \apj, 270, 88
\bibitem[]{} Jura, M. \& Kleinmann, S. G. 1989, \apj, 341, 359
\bibitem[]{} Kennicutt, R. C. 1983, \apj, 272, 54
\bibitem[]{} Kennicutt, R. C., Keel, W. C. \& Blaha, C. A. 1989, \aj, 97,1022
\bibitem[]{} Kennicutt, R. 1998, \apj, in press.
\bibitem[]{} Lehnert, M. D. 1993, dissertation, The Johns Hopkins University
\bibitem[Lehnert \& Heckman 1994]{Lehheck94} Lehnert, M. D. \& Heckman, T. M. 
  1994, \apj, 426, L27 
\bibitem[Lehnert \& Heckman 1995]{} Lehnert, M. D. \& Heckman, T. M. 1995, 
  \apjs, 97, 89
\bibitem[Lehnert \& Heckman 1996a]{} Lehnert, M. D. \& Heckman, T. M. 1996a, 
  \apj, 462, 651
\bibitem[Lehnert \& Heckman 1996b]{} Lehnert, M. D. \& Heckman, T. M. 1996b, 
  \apj, 472, 546
\bibitem[]{} Leitherer, C. \& Heckman, T. M. 1995, \apjs, 96, 9
\bibitem[]{} Lord, et al. 1996, \apj, 465, 703
\bibitem[]{} Marlowe, A. T. et al. 1995, \apj, 438, 563
\bibitem[]{} Martin, C. L. 1997, \apj, 491, 561
\bibitem[]{} McKee, C. F. \& Ostriker, J. P. 1977, 218, 148
\bibitem[]{} McKee, C. F. \& Williams, J. P. 1997, 476, 144
\bibitem[]{} O'Dell, C. R. \& Casta$\tilde{\rm n}$eda, H. O. 1984, \apj, 
  283,158
\bibitem[]{} Osterbrock, D. E. 1989, Astrophysics of Gaseous Nebulae and Active
        Galactic Nuclei(Mill Valley, CA: University Science Books)
\bibitem[Reynolds 1990]{rey90} Reynolds, R.J. 1990, in The Galactic and 
  Extragalactic Background Radiation, ed. S.~Bowyer and C. Leinert 
  (IAU Symp. 139)(Dordrecht: Kluwer), 157
\bibitem[Reynolds 1993]{} Reynolds, R.J. 1993, in Massive Stars: Their Lives 
  in the Interstellar Medium, ed. J. P. Cassinelli and E. B. Churchwell (ASP 
  Conf. Ser. Vol. 35), 338
\bibitem[]{} Savage, B., \& Sembach, K. 1996, ARA\&A, 34, 279
\bibitem[Sokolowski 1993]{sok95} Sokolowski, J. 1993, private communication
\bibitem[]{} Suchkov, A. A., et al. 1996, \apj, 463, 528
\bibitem[]{} Vila-Costas, M. B., \& Edmunds, M. G. 1993, MNRAS, 265, 199 
\bibitem[]{} Vogel, S. N. et al. 1995, \apj, 441, 162
\bibitem[]{} Wang, J. et al. 1997, \apj, 474, 659
\bibitem[]{} Wang, J., Heckman, T.M. \& Lehnert, M.D. 1997, \apj, 491, 114 
  (WHL)
\bibitem[]{} Wang, J. 1998 dissertation, The Johns Hopkins University 
\bibitem[]{} Wang, J., Heckman, T., \& Lehnert, M. 1998, in preparation

\end{thebibliography}
